# Influence of disorder on $H_{c2}$-anisotropy and flux pinning in MgB$_2$

M. Eisterer[*]

Atomic Institute of the Austrian Universities, Stadionallee 2, 1020 Vienna, Austria

The upper critical field and flux pinning in MgB$_2$ single crystals were investigated. The implications of these properties for technical applications are discussed and compared with transport properties of polycrystalline bulk samples and wires. In these untextured materials current percolation is important, especially at high magnetic fields. It is shown that the anisotropy of the upper critical field influences the "irreversibility line" and that the application range of MgB$_2$ is limited by the smallest upper critical field (i.e., for the field direction perpendicular to the boron planes).
Disorder, introduced by irradiation with neutrons, enhances the upper critical field, reduces the anisotropy and drastically changes flux pinning. While the enhanced $H_{c2}$ and the reduced anisotropy generally improve the transport properties of the polycrystalline samples, the contribution of the radiation-induced defects to flux pinning is small compared to the as-grown defect structure (grain boundary pinning).

## 1   Introduction

Since the discovery of MgB$_2$ [1] research not only concentrates on understanding the basic properties of this interesting material, but also on the improvement of parameters that are important for technical applications, such as the critical current and the upper critical field. In the pure material, B$_{c2}$ is smaller than in existing technical superconductors, especially if the field is applied perpendicular to the boron planes. First indications that enhancements of $B_{c2}$ can be obtained by the introduction of disorder were found in thin films [2] and by irradiation experiments [3]. Partial substitution of either Mg [4-6] or B [7,8] offers another interesting possibility not only to introduce disorder but also to modify the electronic structure. Especially carbon doping on the boron sites is promising for the enhancement of the upper critical field [9-11].

It was pointed out [12,13] that grain boundaries do not inhibit the current flow in MgB$_2$, thus making polycrystalline material to an interesting candidate for technical applications. Current percolation plays an important role in these untextured materials [14], especially in magnetic fields [15], since - due to the H$_{c2}$ anisotropy - a magnetic field induces different transport properties, depending on the orientation of a grain's boron plane to the applied field.

In this paper I will survey systematically the influence of disorder introduced by neutron irradiation on the transition temperature, on the upper critical field and its anisotropy and on flux pinning in single crystals and in polycrystalline MgB$_2$ and discuss the implication of these changes on the percolative current transport.

## 2   Neutron induced disorder

Neutron irradiation was made in the TRIGA-MARK-II research Reactor in Vienna. Two different irradiation positions were used: The central irradiation facility with a fast/thermal neutron flux density of $7.6/6.1\times10^{16}$ m$^{-2}$s$^{-1}$ [16] and a position located outside the graphite reflector, where the fast and thermal neutron flux densities are 1.4 and $30\times10^{14}$ m$^{-2}$s$^{-1}$, respectively. While a fast neutron has enough energy to displace an atom from its lattice position by a direct collision, low energy neutrons can induce disorder only by the neutron capture reaction of the $^{10}$B isotope (19.9 % in natural boron). The activated nucleus emits an alpha particle and $^{7}$Li nucleus with a kinetic energy of 1.7 MeV and 1 MeV, respectively. Most of this energy is transferred to the electron system along the range of the reaction products (4.8 and 2.1 μm for the $^{4}$He and the $^{7}$Li nucleus, respectively) and does not produce defects in the crystal lattice. Only a small fraction of the energy loss is due to nuclear reactions, resulting in about 320 dpa per alpha particle generated [3]. Immediate recombination reduces the number of stable defects, but this effect cannot be estimated quantitatively. The reaction cross section for the $^{10}$B(n,α)$^{7}$Li reaction strongly depends on the neutron energy and becomes huge at low neutron energies (e.g. 3837 b at 25.3 meV). All these neutrons are absorbed at the sample's surface, which results in an inhomogeneous defect density. This effect can be suppressed by shielding the sample with cadmium during the irradiation, which absorbs nearly all neutrons with energies below 0.5 eV. In any case the number of primary recoils due to the $^{10}B(n,\alpha)^{7}Li$ reaction is much larger than that of the direct collisions. The size of defects ranges from point defects (~ 0.1 nm) to collision cascades ( ~ 10 nm). Such large defects have been found in TEM images of irradiated single crystals [17]. A typical example is shown in Fig. 1. Investigations on whether these large defects result from the n-α reaction or from the collisions of fast neutrons, are currently under way.

[*] Corresponding author: e-mail: eisterer@ati.ac.at

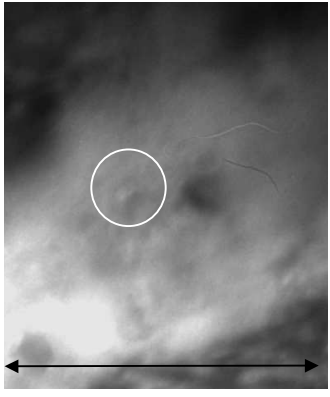 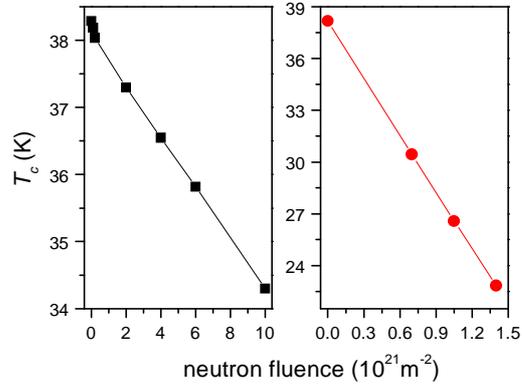

**Fig. 1** Dark-field image of a radiation induced defect.

**Fig. 2** Decrease of transition temperature following fast (left) and thermal (right) neutron irradiation.

## 3 Changes of the reversible properties

Several single crystals [18] were exposed to fast neutrons in the central irradiation facility, the thermal neutrons were shielded with cadmium. The transition temperature was determined by ac susceptibility at an amplitude of 30 µT (linear extrapolation of the steepest slope to zero). After a faster decrease at low fluences (Fig. 2, left), we find a linear decrease of the transition temperature with fast neutron fluence (-3.7 K per $10^{22}$ m$^{-2}$). For comparison another single crystal was sequentially irradiated without cadmium shield outside the graphite reflector, mainly with thermal neutrons. The dependence of $T_c$ on the thermal neutron fluence is also linear with a slope of about –10.9 K per $10^{21}$ m$^{-2}$. This much faster change is a consequence of the larger cross section of the $^{10}B(n,\alpha)^{7}Li$ reaction at thermal neutron energies. The decrease of $T_c$ caused by disorder in MgB$_2$ is commonly explained [19,20] by interband scattering between the π- and σ-bands and is expected to be caused by rather small defects. The linearity of the change down to at least 23 K demonstrates that the transition temperature is a reasonable measure of the disorder introduced by neutrons, independently of the actual irradiation technique (fast-thermal, Cd-shielded etc.).

The disorder also changes the upper critical fields of MgB$_2$, as can be seen in Fig. 3. The data were obtained from SQUID magnetometry by extrapolating the linear decrease of the magnetic moment with increasing temperature (near $B_{c2}$) to zero. We find a monotonic increase of $B_{c2}$ at zero temperature with increasing disorder for the field perpendicular to the boron planes. Within experimental error it cannot be decided, if the sample with the lowest $T_c$ (26.5 K) has a higher $B_{c2}$ than the sample with a transition temperature of 30.4 K, but it can be concluded, that the maximum $B_{c2}$ is obtained in this temperature range. For the other main field direction (parallel to the ab-plane) we initially find an increase of $-\partial B_{c2}/\partial T$ near $T_c$, resulting in higher values at low temperatures.

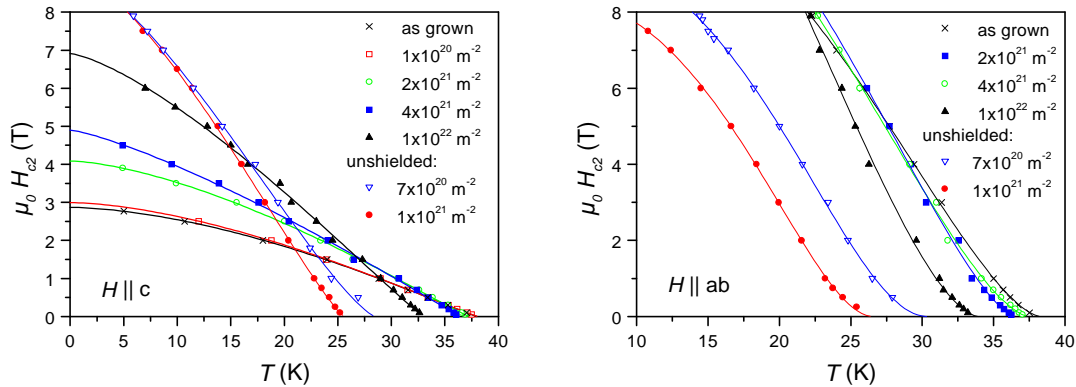

**Fig. 3** Upper critical fields of single crystals perpendicular (left) and parallel (right) to the boron planes after irradiation to various fluences. Lines are a guide for the eyes only.

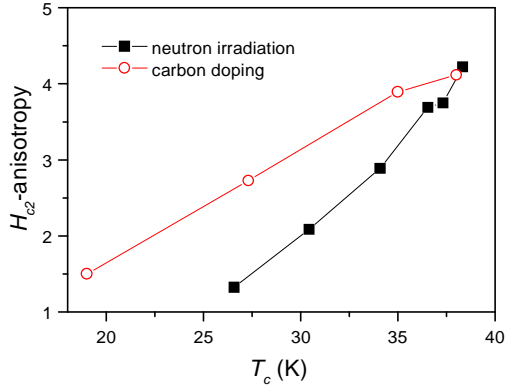 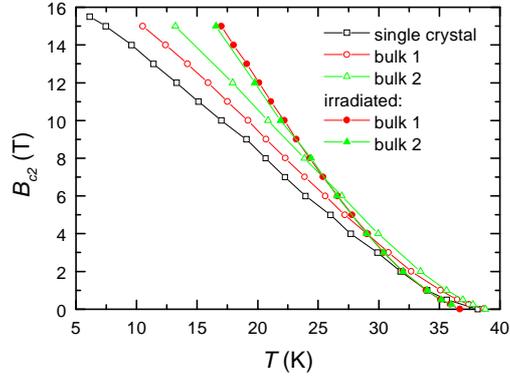

**Fig. 4** Influence of disorder on $H_{c2}$-anisotropy at $T/T_c$=0.6. Data on carbon doped crystals were extracted from [10].

**Fig. 5** Upper critical fields of polycrystalline bulk samples in comparison to a single crystal.

With decreasing $T_c$ the slope and, therefore, $B_{c2}(0)$ starts to decrease again. The generally smaller effect for this field direction leads to a reduction of the $H_{c2}$-anisotropy. In Fig. 4 the anisotropy at $t$=0.6 ($t=T/T_c$) is plotted as a function of the transition temperature. This reduced temperature was chosen in order to avoid extrapolation of the data to low temperatures, which might induce large errors. The anisotropy decreases from 4.2 for the as-grown crystal to about 1.3 for the crystal with $T_c$=26.5 K. The kink at high transition temperatures should not be taken too seriously, because of experimental uncertainties. Note that the slope does not decrease much with decreasing anisotropy, although the lowest value is close to the isotropic case.

Figure 5 compares the upper critical fields (obtained from transport measurements) of a single crystal for $H||ab$ with two different bulk samples, which strongly differ in their normal state resistivity. It was found to be 4.8 µΩcm for sample 1 [21] and 85 µΩcm for sample 2 [13]. Since in polycrystalline materials grains with their boron plane oriented parallel to the applied field become superconducting at first, they determine the upper critical field. The upper critical fields of the bulk samples are slightly higher, but comparable to those of the single crystal. After irradiation, $T_c$ decreases from 38.7 K to 36.7 K in sample 1 and from 38.8 K to 36.5 K in sample 2. Both samples now behave nearly identically with generally higher values than before the irradiation, except at high temperatures. The increase of $B_{c2}$ caused by the introduction of defects is usually explained by a reduction of the mean free path of the charge carriers. This reduces the coherence length and, therefore, enhances $B_{c2}$. In fact, the normal state resistivity at 40 K increases by 11 and 15 µΩcm after neutron irradiation for sample 1 and sample 2, respectively. Also, the higher $B_{c2}$ of the unirradiated sample 2 could result from the mean free path effect, since its resistivity is higher by more than one order of magnitude compared to sample 1. Although there is experimental evidence that a high resistivity favours a large upper critical field [2,22], it is impossible to find a universal relationship between the resistivity and $B_{c2}$. This is a consequence of the two band nature of $MgB_2$, since scattering centres have different effects on the resistivity and $B_{c2}$ depending on which charge carriers (σ or π) they influence most [23]. In polycrystalline samples a contribution of the grain boundaries and current percolation can enhance the resistivity, too [14]. Nevertheless, we find some interesting similarities to data on carbon doped $MgB_2$. In polycrystalline samples an increase of the resistivity by about 10 µΩcm was found at a doping level of 3.8 % [9] compared to the undoped reference sample, accompanied by a decrease of the transition temperature by 3 K (2 and 2.8 K for our samples). $B_{c2}$ of the neutron irradiated samples becomes 15 T at about 17 K, at about 20 K in the carbon doped sample. In Fig. 6, a carbon doped single crystal [24] with a $T_c$ of 35.7 K is compared to a fast neutron irradiated crystal with a slightly smaller $T_c$ (34.3). Apart from the shift of the transition temperature, both samples are quite similiar. Also the general change of the upper critical fields for both field directions is comparable [10,11]. While in both cases the maximum reachable $B_{c2}^{\perp}$ seems to be around 10 T, the maximum $B_{c2}^{||}$ is higher in the carbon doped samples. This leads to a smaller decrease of the anisotropy with decreasing transition temperature (Fig. 4). The generally similar behaviour and the differences between carbon doping and neutron irradiation should help to clarify the influence of disorder in comparison to changes of the charge carrier concentration and of the topology of the Fermi-surface, which are important for carbon doped samples [10,25].

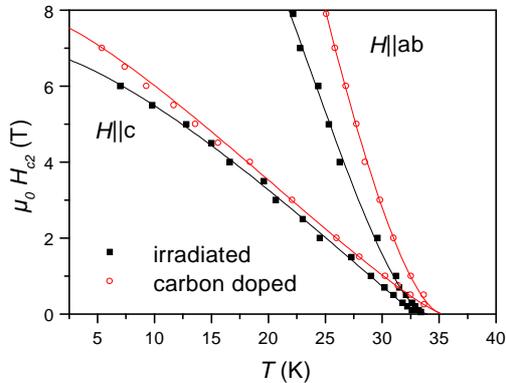

**Fig. 6** Comparison of the upper critical fields of an irradiated and of a carbon doped single crystal.

While $B_{c2}(T)$ of our single crystal and our bulk sample 1 are similar, the resistive transition differs strongly. Whereas in zero field the transition width of both sample is only 0.3 K, indicating excellent homogeneity of both samples, the transition broadens significantly more in magnetic fields for the polycrystalline sample (Fig. 7), although the applied current density was much smaller in the latter case (single crystal: $10^5$ Am$^{-2}$, bulk: $10^4$ Am$^{-2}$). This additional broadening was identified to be a result of the $H_{c2}$-anisotropy [26,15]. For a continuous superconducting current path a certain fraction $p_c$ of superconducting grains is necessary, which depends on the number of connections between the grains and which is expected to be between 0.17 and 0.3. In a polycrystalline untextured material the grains are randomly oriented and, due to anisotropy, the upper critical field of each grain depends on the angle between its boron plane and the applied field. At a fixed field B the transition starts at a temperature T at which $B=B_{c2}^{\parallel}(T)$. With decreasing temperature more grains become superconducting until their fraction is $p_c$, where the resistivity disappears. This leads to a transition width caused by anisotropy [15]:

$$\Delta T = \frac{\left(1-\sqrt{(\gamma^2-1)p_c^2+1}\right)}{\frac{\partial B_{c2}}{\partial T}} B \qquad (1)$$

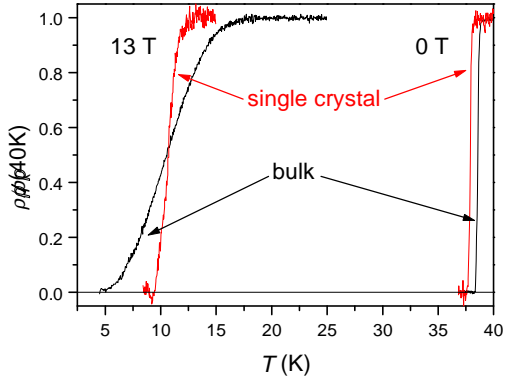 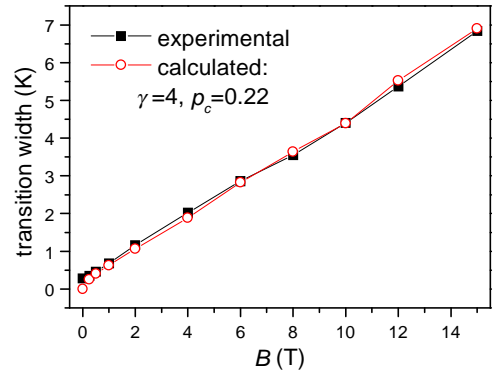

**Fig. 7** Resistive transitions of a single crystal and of a polycrystalline bulk sample at 0 T and at 13 T.

**Fig. 8** Transition width as derived from resistive measurements compared to a theoretical model.

In Fig. 8 the experimental data of sample 1 are compared with the prediction of Eq. 1. The transition width was chosen as the difference in temperature between 95 % and 5 % of the normal state resistivity at 40 K. For the calculation I used a temperature independent $\gamma$ of 4, $p_c$=0.22 and $\partial B_{c2}/\partial T$ from the experimental data. The agreement between the theoretical behaviour and the experimental data is excellent, especially the proportionality between the transition width and the applied field is reproduced in the experiment. Minor deviations from this linear dependence are due to the temperature dependence of $\partial B_{c2}/\partial T$ and of anisotropy. Note that the chosen parameter are only examples and cannot be obtained from the transition width, since $\Delta T$ depends only on the product $(\gamma^2-1)p_c^2$. The magnetic fields, at which the resistivity of the bulk samples disappears, are plotted in Fig. 9 (left). Alike $B_{c2}$, sample 1 has lower values before the irradiation, but the differences become small after irradiation. The disorder induced shift of the zero resistance fields is larger than the shift of $B_{c2}$, e.g. at 12 T the zero resistivity temperature is shifted by 7.7 K and by 3.7 K, while $T_c(B)$ is only shifted by 4.2 and by 1.8 K for sample 1 and sample 2, respectively. This larger effect can be explained by a reduction of anisotropy, as observed in single crystals. The zero resistance fields for two different wires are plotted in the right hand panel of Fig. 9. The copper sheathed wire prepared by an in-situ process [27] has rather small values. This can be explained by the ductility of copper, which does not allow for an efficient compression of the superconductor or its precursor powders, leading to a poor density of MgB$_2$, thus enhancing $p_c$.

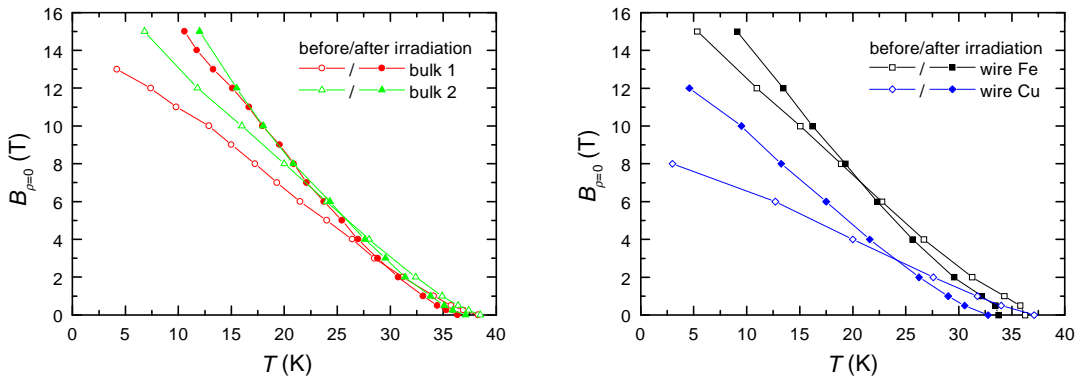

**Fig. 9** "Irreversibility" fields of bulk samples (left) and wires (right) before and after irradiation.

It was found from magnetization measurements, that also $B_{c2}$ of this wire was comparatively small. Unfortunately, the onset of the resistive transition cannot be evaluated with regard to $B_{c2}$ in wires with a conducting sheath, since its conductivity masks the real onset of the transition. Therefore no high field data are available. The zero resistivity fields of the iron sheathed wire nearly reach the high values of the bulk sample 2, but the shift after the irradiation is only small.

## 4 Changes in flux pinning

Magnetization loops of single crystals of $MgB_2$ are reversible over a wide temperature and field range. The introduction of effective pinning centres changes this behaviour, as can be seen in Fig. 10. All measurements were made with the field parallel to the c-axis and $J_c$ was calculated from the irreversible magnetic moment using the Bean model. The reduced temperature was chosen to be 0.13 (~ 5 K), only the two crystals, which were irradiated without cadmium shield, were measured at slightly higher reduced temperatures (0.16 and 0.19), corresponding to 5 K. The as-grown crystal has finite critical currents only at low fields, no peak near the upper critical field can be detected within resolution. At low fluences ($5\times10^{19}$ m$^{-2}$ - $2\times10^{20}$ m$^{-2}$) we find significant radiation effects only at low fields and near the upper critical field, where a peak appears. At intermediate fields the crystal remains reversible. With increasing neutron fluence, the peak grows and the onset shifts to lower reduced fields. Then the peak starts to decrease again, while the onset still shifts to lower reduced fields ($10^{22}$ m$^{-2}$). With increasing disorder, the onset shifts again to higher fields, which cannot be explained only by the higher reduced temperature, but by the decrease of the condensation energy. Also, differences caused by the different irradiation conditions cannot be excluded. The fluence dependence of the emerging peak was explained in [17] by an order-disorder transition of the flux line lattice [28-30]. The field $B_{od}$, at which the order-disorder transition occurs, is plotted as a function of neutron fluence in Fig. 11 and compared to the theoretical of Ref. [30]. We find good agreement, given the simplifications of the model and possible errors in the estimation of the anisotropy at 0 K, although the decrease of the reduced $B_{od}$ with increasing disorder is slower than predicted. $B_{od}$ was defined by the kink in the magnetization curve, but evaluations of the onset or of the position of the maximum of the peak, lead qualitatively to the same results.

The critical current densities of polycrystalline samples are completely different from those in single crystals (Fig. 12). The data for the single crystals were obtained from magnetization loops at 5 K. For the field parallel to the boron planes, $J_c$ represents an ill defined average over currents flowing parallel and perpendicular to the boron planes, the absolute values should not be taken too seriously. I only want to point out the fundamental difference of the pinning properties in single crystals and in polycrystalline materials. The obvious difference in the defect structure are the grain boundaries, which has turned out to represent efficient pinning centres in metallic superconductors and which are also natural candidates for being the dominant pinning centres in polycrystalline $MgB_2$. The critical current densities of the wires were obtained from transport measurements in liquid helium with an 0.1 μΩcm$^{-1}$ criterion, and refer to the superconducting cross section. $J_c$ of the bulk sample 2 was derived from ac susceptibility measurements at 5 K with ac amplitudes between 0.5 and 10 mT at 9 Hz [31]. At low fields $J_c$ of the copper sheathed wire and of the bulk sample is nearly identical, but $J_c$ of the wire decreases much faster with increasing field, a consequence of its low irreversibility (zero resistivity) field. The iron sheathed wire has by far the highest critical currents in the whole experimentally accessible field range and a similar field dependence as the bulk sample. The irradiation enhances $J_c$ of all samples, especially at high fields. The lines in Fig. 12 are fits to a model based on percolation theory [15]. It assumes that $J_c$ of each grain depends on the angle between its boron plane and the applied field (same as $B_{c2}$). The effective cross section for a given current density is calculated by means of percolation theory, depending on the fraction of grains which actually can carry that current density. The model contains three parameters: $J_0$ as a measure of the pinning strength, the $H_{c2}$ anisotropy $\gamma$ and the percolation threshold $p_c$. These parameters were calculated by a fitting procedure and the pinning strength was found to increase only little, $J_0$ changes from 3.2 to 4 and from 3.3 to 3.2×10$^9$ Am$^{-2}$ in the bulk and in the wire, respectively. This can be understood by the small pinning force of the radiation induced defects compared to that of the large grain boundaries. The anisotropy $\gamma$ decreases from 4.4 to 2.8 in the bulk and from 4.4 to 2.6 in the wire. The main effect of the irradiation is this reduction of anisotropy together with the enhancement of the upper critical field. The decrease of anisotropy is in reasonable agreement with our findings for single crystals (Fig. 4). The data on the iron sheathed wire could not be fitted appropriately due to the lack of low field data and of an estimation for the upper critical field $B_{c2}^{\parallel}$.

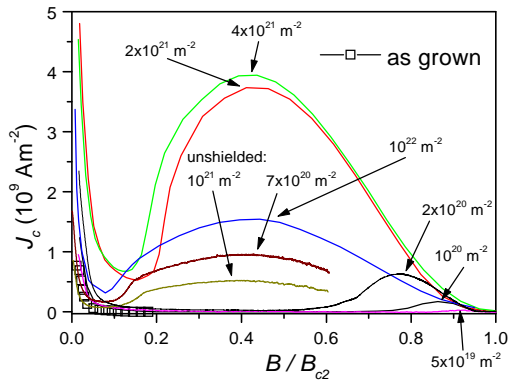

**Fig. 10** Critical current densities at ~5 K (see text) of various neutron irradiated single crystals ($H$||c).

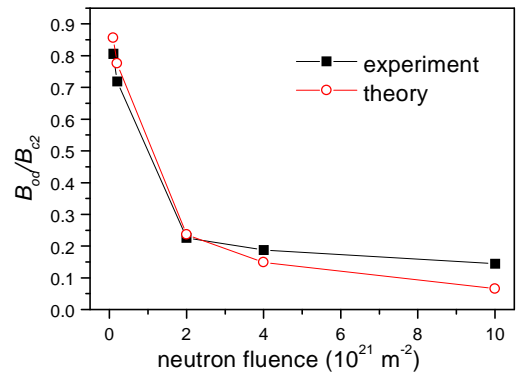

**Fig. 11** Fluence dependence of the order-disorder transition.

The influence of variations in $\gamma$ or in $p_c$ on the field dependence of $J_c$ is demonstrated in Fig. 13. The magnetic field is normalized by $H_{c2}^{\perp}$, $H_{c2}^{\parallel}$ is equal to gamma in that representation. The open squares represent the fitted curve of the unirradiated bulk sample 2, $p_c$ is then changed from 0.21 to 0.25 and to 0.17. In real samples a change in $p_c$ can be obtained by modifying the density of superconductor and the size and shape of its grains. Although the change of $p_c$ shifts the curve at high fields, resulting in different irreversibility fields defined by very low current densities, it does nearly not change $J_c$ for high current densities, which are necessary for technical applications. Note that all $J_c$'s drop to below 5×10$^7$ Am$^{-2}$ very close to $H_{c2}^{\perp}$ (=1). This is also true, when the anisotropy is changed. Since $H_{c2}^{\perp}$ is fixed, a reduced anisotropy implies a lower $H_{c2}^{\parallel}$ in this representation. A change of the pinning strength $J_0$ shifts the whole curve upwards (on the logarithmic scale) and does not influence the field dependence at all. Therefore, the field at which $J_c$ becomes 5×10$^7$ Am$^{-2}$ also changes, but the shift is not very large, if $J_0$ is not changed drastically. We conclude, that mainly $H_{c2}^{\perp}$ limits the field range of applicability of polycrystalline MgB$_2$.

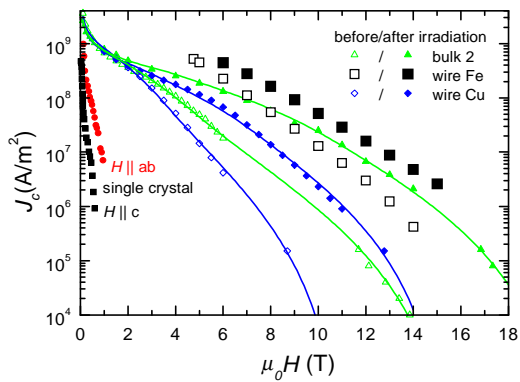 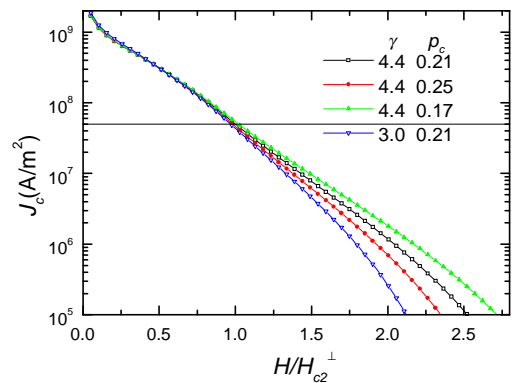

**Fig. 12** Critical current densities of a single crystal, of a bulk sample and of two wires.

**Fig. 13** Influence of the $H_{c2}$-anisotropy and of the percolation threshold on the critical current densities.

## 5 Conclusions

The introduction of disorder by neutron irradiation leads to a decrease of the transition temperature, to an increase of the upper critical fields for both main field directions and to a reduction of anisotropy. These changes are a consequence of enhanced scattering of the charge carriers, which is supported by the observed increase of the normal state resistivity. Similarities to the introduction of disorder by carbon doping indicate that also in that case scattering is most important and the observed differences might be a result of the change of the charge carrier density and of the topology of the Fermi surface after carbon doping.

In polycrystalline samples, grain boundaries were identified as the dominant pinning centres. Therefore, the introduction of additional pinning centres did not change pinning strongly, in contrast to the situation in single crystals, where the additional defects cause a transition from a highly ordered vortex lattice to a disordered (glassy) state. It was shown that in polycrystalline samples mainly the lowest upper critical field (i.e. for $H\|c$) limits the useful field range for technical applications.

**Acknowledgements** I wish to thank Professor Harald W. Weber for useful discussions and for critically reading the manuscript. Special thanks go to Martin Zehetmayer for providing me with some of his data and for fruitful discussions. I also wish to thank Christian Krutzler, Rainer Prokopec and Regina Müller for taking some of the measurements. I am grateful to Balaji Birajdar and Oliver Eibl (Universität Tübingen) for the TEM investigations. Samples were provided by Sergei Kazakov and Janusz Karpinski (ETH Zürich), by Tsuyoshi Tajima (LANL), by Sonja Schlachter and Wilfried

Goldacker (FZ Karlsruhe) and by N. Hari Babu, Bartek Glowacki and David Cardwell (IRC Cambridge), which is gratefully acknowledged.